\documentstyle[12pt]{article}

\newcommand{\EQ}{\begin{equation}}
\newcommand{\EN}{\end{equation}}

\begin{document}

\topmargin 0pt
\oddsidemargin 5mm
\newcommand{\NP}[1]{Nucl.\ Phys.\ {\bf #1}}
\newcommand{\PL}[1]{Phys.\ Lett.\ {\bf #1}}
\newcommand{\NC}[1]{Nuovo Cimento {\bf #1}}
\newcommand{\CMP}[1]{Comm.\ Math.\ Phys.\ {\bf #1}}
\newcommand{\PR}[1]{Phys.\ Rev.\ {\bf #1}}
\newcommand{\PRL}[1]{Phys.\ Rev.\ Lett.\ {\bf #1}}
\newcommand{\MPL}[1]{Mod.\ Phys.\ Lett.\ {\bf #1}}
\newcommand{\JETP}[1]{Sov.\ Phys.\ JETP {\bf #1}}
\newcommand{\TMP}[1]{Teor.\ Mat.\ Fiz.\ {\bf #1}}

\renewcommand{\thefootnote}{\fnsymbol{footnote}}

\newpage
\setcounter{page}{0}
\begin{titlepage}
\begin{flushright}
SISSA-EP-85
\end{flushright}
\vspace{0.5cm}
\begin{center}
{\large  Exact resonance A-D-E S-matrices and their
renormalization group trajectories
} \\
\vspace{1cm}
\vspace{1cm}
{\large  M\'arcio Jos\'e  Martins
\footnote{on leave from Departamento de Fisica, Universidade Federal de
S.Carlos, C.P. 676 - S.Carlos 13560, Brazil}
\footnote{martins@itssissa.bitnet}} \\
\vspace{1cm}
{\em International School for Advanced Studies\\
34014, Strada Costiera 11, Trieste,
 Italy}\\
\end{center}
\vspace{1.2cm}

\begin{abstract}
We introduce the A-D-E resonance factorized models as an appropriate
analytical continuation of the Toda S-matrices to the complex values
of their coupling constant. An investigation of the associated Casimir
energy, via the thermodynamic Bethe ansatz, reveals a rich pattern
of renormalization group trajectories interpolating between the central
charges of the $G_1 \otimes G_k / G_{k+1}$ GKO coset models. We have
also constructed the simplest resonance factorized model satisfying the
``$\phi^3$''-property. From this resonance scattering, we
predict new flows in non-unitary minimal models.
\end{abstract}
\vspace{.5cm}
\vspace{.5cm}
\centerline{May 1992}
\vspace{.3cm}
\end{titlepage}

\renewcommand{\thefootnote}{\arabic{footnote}}
\setcounter{footnote}{0}

\newpage
\section{Introduction}
Conformal invariance has played an important role in our understanding
of the properties of two-dimensional systems at the critical
point \cite{BPZ,CAR}. Away from the critical point, the corresponding
field theory can be considered as a conformal model perturbed by
an appropriated relevant operator. The important insight added by
A. Zamolodchikov \cite{BZA1} was that certain perturbations may
possess infinitely many non-trivial conserved charges, hence integrable
even in the off-critical regime. This observation allowed him to
construct the on-mass-shell solution, i.e. the mass spectrum and the exact
S-matrices, of the Ising model in a magnetic field \cite{BZA2}. In this
context, Al. Zamolodchikov \cite{ZAM1} has pointed out that by applying the
thermodynamic Bethe ansatz~(TBA) \cite{YY} one is able to recover
the ultraviolet properties of the ground state from the respective
S-matrices. In this sense, the TBA makes an interesting bridge between
the properties of the infrared and the ultraviolet regimes. In principle,
this technique can be applied to any factorized scattering theory, even
for those that a priori lack a background conformal field theory. An
a posteriori analysis of the behaviour of the finite-size corrections
to the ground state may reveal amazing features not previously known.
Recently, one
example has been put forward by Al. Zamolodchikov \cite{ZAM2} in the case of
a resonance S-matrix consisting  of a single particle scattering through
the following amplitude,
\EQ
S(\theta,\theta_0)= \frac{\sinh(\theta)-i \cosh(\theta_0)}
{\sinh(\theta)+i \cosh(\theta_0)}
\EN
where $\theta_0$ is the resonance parameter.

Let us briefly summarize Al. Zamolodchikov's results \cite{ZAM2}. Although
the ultraviolet limit associated with
this S-matrix is always governed by a theory
with central charge $c=1$, at intermediate distances~(for $\theta_0 >>1$)
the pattern of the renormalization group~(RG) trajectories is surprisingly
rich. The RG trajectories evolve to the infrared regime by first
interpolating between the central charges $c_p=1-6/p(p+1)$ of the minimal
models. Each fixed point $c_p$ takes approximately $\frac{\theta_0}{2}$
of the RG ``time'' before crossing over to the next lower critical point
$c_{p-1}$. At this point we recall that such crossover phenomena
can be induced by the perturbation of the least relevant field operator
$\phi_{1,3}$ of the minimal models \cite{BZAM3,LUCA,KMS}. Recently, it
has also been argued in ref. \cite{MI1} that the field $\phi_{1,3}$
can produce crossover behaviour even in non-unitary minimal models.

The purpose of this paper \footnote{A brief account of this paper has recently
been  presented in ref.\cite{MA}}
is to present and analyze a large class
of resonance factorized scattering theories based on the simply
laced Lie algebra G. These S-matrices will define a family of one
parameter field theories $G(\theta_0)$, depending on the resonance parameter
$\theta_0$. We point out a connection between these resonance S-matrices
and those from the A-D-E Toda field theories \cite{AFZ,CRMU,BC,DV,MU} through
an appropriate analytical continuation of the Toda coupling constant to
complex values. This allows us to express the resonance S-matrices amplitudes
in a rather simple and useful form.
We study the Casimir energy $E_0(R,\theta_0)$ at
moderate distances R via the TBA approach. The behaviour of the
associated RG trajectories
are then related
to flows \cite{FLO} of certain deformed GKO coset models
\cite {GKO}. Finally, we propose a simplest resonance factorized model
satisfying the ``$\phi^3$''-property. We argue that the respective
RG trajectories can be related to new flows in the non-unitary
minimal models.

The paper is organized as follows. In sect.2 we introduce the A-D-E
resonance models and write down the respective TBA equations. Sect.3
is devoted to our numerical checks of the behaviour of the ground
state energy $E_0(R,\theta_0)$ for finite values of the volume R. Sect.4
describes the ``$\phi^3$''-resonance model and our arguments to predict
new flows in non-unitary minimal models. Sect.5 contains a discussion
of our results. In appendix A we summarize some useful relations in
order to compute the logarithmic corrections to the Casimir energy when
$R \rightarrow 0$.

\section{The A-D-E resonance S-matrices and the TBA equations}
We start this section by describing the basic properties of the
resonance scattering that we have constructed. We assume that the
S-matrices are diagonal and factorizable, i.e., there is no particle
production. In this case, the only constraints are due to the unitarity
and crossing conditions of the two-body amplitude
$S_{a,b}$, \footnote{In our case the antiparticle appears as a pole
in the particle-particle scattering amplitude,
and therefore the reflection amplitude
is null in order to satisfy factorizability \cite{KS}.}
\EQ
S_{a,b}(\theta) S_{a,b}(-\theta)=1,~~ S_{a,b}(\theta)=
S_{b,\bar{a}}(i \pi -\theta)
\EN
where $\bar{a}$ stands for an antiparticle.

We also assume that the resonance amplitude $S_{a,b}(\theta,\theta_0)$ can
be written as a product of a background amplitude
$S_{a,b}^{mim}(\theta)$ and
the resonance part $S_{a,b}^{res}(\theta,\theta_0)$, namely
$S_{a,b}(\theta,\theta_0)=S_{a,b}^{min}(\theta)
S_{a,b}^{res}(\theta,\theta_0)$.
The physical meaning of this hypothesis is that the background and
the resonance interactions appear in different spatial regions, and
therefore the phase-shifts are simply added. More complicated cases
will involve correlations that will not be considered in this paper.
However, as we  have already stated, even this simple ansatz will lead us
to a rather rich pattern of RG trajectories appearing in deformed conformal
field theory.

In order to introduce the A-D-E structure we assume that the particle spectrum
and the background S-matrices, $S_{a,b}^{min}(\theta)$, are given by
the so-called minimal solution of Eq.(2). For details and respective
properties see
ref. \cite{AFZ,CRMU,BC,DV,MU}. The simplest solution of Eq.(2) to
the resonance amplitude $S_{a,b}^{res}(\theta,\theta_0)$, invariant by
the symmetry $\theta_0 \rightarrow -\theta_0$, appears in the case of
the $A_{N-1}$ Lie algebra. The S-matrix for the fundamental particle
is given by
\EQ
S_{1,1}^{res}(\theta,\theta_0)=
\frac{\sinh\frac{1}{2}(\theta-\theta_0-i\frac{\pi}{N})}
{\sinh\frac{1}{2}(\theta-\theta_0+i\frac{\pi}{N})}
\frac{\sinh\frac{1}{2}(\theta+\theta_0-i\frac{\pi}{N})}
{\sinh\frac{1}{2}(\theta+\theta_0+i\frac{\pi}{N})} ,
\EN
where the resonance poles are located at $\theta=\pm \theta_0
-i\frac{\pi}{N}$. The other amplitudes can be  computed  by
a straightforward bootstrap. For N=2, we recover Al. Zamolodchikov's
S-matrix, Eq.(1).

Similar expressions can be found for the D and E groups, by consistently
solving Eq.(2) in analogous manner as
pursued in refs. \cite{AFZ,CRMU,BC,DV,MU}. However, we
prefer to take a short route.
First, we notice that the amplitude $S_{1,1}^{res}(\theta,\theta_0)$
above is related in a simple way to the so-called Z-factors of the
$A_{N-1}$ Toda field theory \cite{AFZ}. These factors encode the
dependence of the coupling constant $\alpha$ present in the Lagrangean,
through a function $b(\alpha)$ \cite{AFZ}. Then, by simply taking
$b(\alpha)=\pm i \theta_0 +\frac{\pi}{N}$ in the Toda
$Z_{1,1}(\theta,b(\alpha))$-factor we
recover the amplitude of Eq.(3) \footnote{The case N=2 was first noticed
by Al. Zamolodchikov \cite{ZAM2} in the context with the sinh-Gordon model.}.
In the general case, we have verified that the resonance S-matrix satisfying
the properties mentioned above is given by \footnote{Here we have followed the
notation of ref. \cite{AFZ}. There are other similar
notations in the
literature \cite{CRMU,BC,DV,MU}. For example, in the case of ref. \cite{BC} ,by
taking $B(\alpha) =1 \pm i \theta_0 \frac{\pi}{h}$
we achieve the same results.},
\EQ
S_{a,b}^{res}(\theta,\theta_0)= Z_{a,b}(\theta,b(\alpha)=
\pm i \theta_0 +\frac{\pi}{h})
\EN
where $Z_{a,b}(\theta,b(\alpha))$ are the Z-factors appearing in the
A-D-E Toda field theory \cite{AFZ,CRMU,BC,DV,MU} and $h$ is the Coxeter
number.

Using Eq.(4) we can write down a closed expression for all the
$S_{a,b}(\theta,\theta_0)$ amplitudes. First it is convenient to define
the functions,
\EQ
\psi_{a,b}(\theta,\theta_0)=-i \frac{d}{d \theta} \log S_{a,b}(\theta,\theta_0)
\EN
which will also be important in the TBA equations. In terms of its Fourier
component $\tilde{\psi}_{a,b}(k,\theta_0)$ defined by,
\EQ
\tilde{\psi}_{a,b}(k,\theta_0)= \int_{-\infty}^{\infty} e^{i k \theta}
\psi_{a,b}(\theta,\theta_0)
\EN
we find the remarkable matrix identity,
\EQ
{\left [ \delta_{a,b} -\frac{\tilde{\psi}_{a,b}(k,\theta_0)}{2 \pi}
\right ]}^{-1}= \frac{\delta_{a,b} \cosh[\frac{\pi k}{h}] -l_{a,b}/2}
{\cosh[\frac{\pi k}{h}] -\cos(k \theta_0)}
\EN
where $l_{a,b}=2-C_{a,b}$ is the incident matrix  of the
G=A,D,E Lie algebra. Finally, the amplitude $S_{a,b}(\theta,\theta_0)$
can be written in terms of a standard integral representation,
\EQ
S_{a,b}(\theta,\theta_0)=exp \left ( \frac{i}{\pi}
\int_{0}^{\infty} \tilde{\psi}_{a,b}(k,\theta_0) \frac{\sin(\theta k)}{k}
dk \right )
\EN

We now start to discuss the TBA equations and their properties. The TBA
approach allows one \cite{ZAM1} to compute the Casimir energy
$E_0(R,\theta_0)$ on a radius of length R. The ground state energy is
written in terms of pseudoenergies $\epsilon_a(\theta)$, and for
diagonal S-matrices there is one such function $\epsilon_a(\theta)$
for each stable mass $m_a$ of the spectrum. The $E_0(R,\theta)$ is
given by,
\EQ
E_0(R,\theta_0) = -\frac{1}{2 \pi}
\sum_{a=1}^{N} m_a \int_{-\infty}^{\infty} d\theta cosh(\theta)
L_a(\theta)
\EN
where $L_a(\theta)=\ln(1+e^{-\epsilon_a(\theta)})$, and N is the
number of particles.

The volume dependence is encoded by a set of integral equations for
the pseudoenergies $\epsilon_a(\theta)$, depending on the kernel
$\psi_{a,b}(\theta,\theta_0)$ as follows,
\EQ
\epsilon_a(\theta) + \frac{1}{2\pi} \sum_{b=1}^{N}
\int_{-\infty}^{\infty} d\theta'
\psi_{a,b}(\theta-\theta',\theta_0) L_b(\theta') =
\nu_a(\theta)~~~,a=1,2,...,N
\EN
where $\nu_a(\theta)=m_a R \cosh(\theta)$ and in our case
$m_a$ are the mass gaps of the A-D-E spectrum \cite{AFZ,CRMU,BC,DV,MU}. It
is possible to rewrite Eq.(10)
in a more suggestive way, adopting an approach similar
to that of ref.
\cite{ZAM3}. Using Eq.(7) and the following identity \cite{ZAM3},
\EQ
\nu_a(\theta+\frac{i \pi}{h})+\nu_a(\theta-\frac{i \pi}{h}) =\sum_{b=1}^{N}
l_{a,b} \nu_b(\theta)
\EN
we then obtain a set of functional equations for the functions
$Y_a(\theta)=e^{\epsilon_a(\theta)}$,
\EQ
Y_a(\theta +\frac{i \pi}{h}) Y_a(\theta-\frac{i \pi}{h})=
\prod_{b \in G} {\left [1+Y_b(\theta) \right ]}^{l_{a,b}}
{ \left [1+Y_a^{-1}(\theta+\theta_0) \right ] }^{-1}
{ \left [1+Y_a^{-1}(\theta-\theta_0) \right ] }^{-1}
\EN

Motived by Al. Zamolodchikov's discussion of the $A_1$ \cite{ZAM2} it is
possible to rewrite Eq.(12)  as an interesting discrete equation if
we restrict the rapidities $\theta$ on the lattice variables $\theta_{m,n}=
\Theta+m \theta_0 +\frac{i \pi}{h} n$. Using the
identification $Y_a^{m,n}=Y_a (\theta_{m,n})$ we
have,
\EQ
Y_a^{m,n+1} Y_a^{m,n-1}= \prod_{b \in G}
{\left [ 1+Y_b^{m,n} \right ]}^{l_{a,b}}
\prod_{c \in A}
{\left [ 1+\frac{1}{Y_a^{c,n}} \right ]}^{-l_{m,c}}
\EN
where b is an index characterizing the nodes of the Dynkin diagram
of G=A,D,E Lie algebra and c is the same index for the A Lie algebra.

Eq.(13) is defined on the product space of the Lie algebras G~(A,D,E) and A,
namely $G \otimes A$. Eq.(13) also assumes a form similar to
those obtained from certain perturbed GKO
coset models \cite{ZAMT,TBA}. It is worth mentioning
that a similar equation can also be defined in the dual space
$A \otimes G$, provide that we take $Y_a \rightarrow Y_a^{-1}$ in
Eq.(12). Therefore they are self-dual when $G \equiv A$.
In this sense these two pairs of equations are richer than the
TBA equation that we start with, Eq.(10). However,
the TBA equations can be easily recovered from Eq.(12) by specifying the
initial
conditions of the functions $\nu_a(\theta)$.

In the next section, we show that even the simple guess for the energies
nodes $\nu_a$, i.e. $\nu_a(\theta)=m_a R \cosh(\theta)$, will generate a rather
interesting pattern of RG trajectories. It turns out that these
trajectories resemble those of unstable particles decaying to intermediate
states until reaching the vacuum. At each intermediate step the
particle has a life-time of $\frac{\theta}{2}$~(the RG ``time'') before
decaying again to the next state.

\section{ The numerical work}
We begin by analyzing the ultraviolet behaviour of Eqs.(9,10). Standard
manipulations of the TBA equations \cite{ZAM1} give us the following
behaviour of the ground state energy,
\EQ
E_0(R,\theta_0)  \simeq - \frac{ \pi r}{6 R}, R \rightarrow 0
\EN
where r is the rank of the respective Lie algebra.

{}From Eq.(14)
we conclude that the background conformal field theory has central
charge $c=r$ \cite{BCN}. In appendix A we discuss the next to
the leading order in R corrections to $E_0(R,\theta_0)$. The first
correction is logarithmic in R , in contrast with the fractional
power behaviour of the typical non-resonance factorized theories \cite{ZAM3}.
In terms of the function $c(R,\theta_0)=-\frac{6 R}{ \pi} E_0(R,\theta_0)$
our result reads
\EQ
c(R,\theta_0)= r-\frac{3 (\theta^2+\frac{\pi^2}{h^2})}
{X^2} \sum_{a,b} C_{a,b}^{-1}
\EN
where $X=\ln(\frac{m_1 R}{2})$.

The behaviour of the function $c(R,\theta_0)$ becomes particularly
interesting at intermediate lengths of R. In this case one needs to
solve numerically the integral equations (10)  using, e.g., standard
iterative procedures. We start by analyzing the simplest case of
our proposed resonance S-matrices, namely the $A_2$ theory \footnote{ A similar
study has been performed by Al. Zamolodchikov in the case of $A_1$
\cite{ZAM1}.}. In fig. 1(a,b,c,d) we show the behaviour of the function
$c(R,\theta_0)$ for $\theta_0=0,10,20,40$. $c(R,0)$
behaves as a smooth function interpolating from the ultraviolet regime,
$c(R \rightarrow 0,0)=2$, to the infrared region. On the contrary, by
increasing $\theta_0$, it is clear that $c(R,\theta_0)$ starts to
form plateaux around the following fixed points, $c_p=2(1-12/p(p+1)),
p=4,5,6,...$ . These critical points correspond to the conformal field
theories possessing a $Z(3)$-symmetry \cite{WAL}. Since the numbers
of plateaux increase with $\theta_0$ one expects that when $\theta_0
\rightarrow \infty$, all the fixed points will be visited by the RG
trajectories until finally reaching the infrared region.

It is also fruitful to examine the same pattern
in terms of the $\beta$-function along
the RG trajectory. In this case, the plateaux will correspond
to zeros of the beta function. Borrowing
Al.Zamolodchikov's definition \cite{ZAM1}
of the $\beta$-function, we have
\EQ
\beta(g)=-\frac{\partial}{\partial X}c(R,\theta_0),~~~~g=r-c(R,\theta_0)
\EN
where the ``coupling constant'' g has been normalized to zero in the
ultraviolet regime.

In Fig. 2(a,b,c) we show $\beta(g)$ for $\theta_0=10,20,40$. One can
notice that the zeros of $\beta(g)$ are precisely at the points
$g=24/p(p+1), p=4,5,6..$. By increasing $\theta_0$, more zeros are
going to appear in $\beta(g)$, however keeping the general shape of
$\beta(g)$ for those
that have already been formed. We stress that our findings are very much
in agreement with Al. Zamolodchikov's results for the $A_1$ case. In general,
for the proposed A-D-E scattering theories, the function $c(R,\theta_0)$
will form plateaux around the values of the central charge $c_p^r$ of the
$G_1 \otimes G_{p-h}/ G_{p-h+1}$ GKO \cite{GKO} coset construction, namely
$c_p^r=r(1-h(h+1)/p(p+1)),p=h+1,h+2,...$ . In Fig. 3(a) we illustrate this
pattern for the first plateaux in the
models $A_2, A_3, D_4$, which share some common central charges. In Fig. 3(b)
we
plot the respective $\beta(g)$ function.

What we can learn from this numerical computation is as follows. Each
time that $X \simeq -(p-h)\frac{\theta_0}{2}$ the function $c(R,\theta_0)$
crosses over from its value $c_p^r$ to the next~(up) value $c_{p+1}^r $
\footnote{ From our figures 1(c),1(d),3(a) we notice that the
``RG time'' $\frac{\theta_0}{2}$ accounts for the plateau and for the finite
size corrections, and both are model dependent.
For example, in the case of
the $A_{N-1}$ a rough estimate of the plateau's length is $\frac{\theta_0}{N}$.
}.
This indicates that at these values of X the TBA equations may present
a rather special behaviour. Indeed, the linearization of these equations
around $X \simeq -(p-h) \frac{\theta_0}{2}$~(or better the functional Eq.(12))
gives us a set of $\theta_0$-independent equations \footnote{ One should notice
that, e.g., the kernel $\psi_{1,1}(\theta,\theta_0)$ has deep peaks
at $\theta=\pm \theta_0,0$, appearing as the interacting
(A Lie algebra structure) and
the diagonal (G Lie algebra structure)
of the linearized TBA equations.}, describing
the RG flows in the $G_1 \otimes G_{p-h}/G_{p-h+1}$ coset models perturbed
{}~(positive perturbation) by an operator $\Phi_p$ with conformal weight
$\Delta_{\Phi_p}=1-h/(p+1)$ \cite{TBA,ZAMT}. On the other hand, the infrared
finite size corrections will be dominated by the dual field $\tilde{\Phi}_p$
with conformal dimension $\Delta_{\tilde{\Phi}_p}=1+h/(p-1)$ \footnote{
For $p=h+1$ this field is not present in the Kac table and
it is replaced by the spinless combination $T \bar{T}$ of the stress
energy tensor $T$ .}. Therefore,
for each plateau ``p'' the ultraviolet behaviour is governed by
$\Phi_{p}$ and finally attracted to the infrared region by $\tilde{\Phi}_p$.
It is tempting to propose that such behaviour may be reproduced if one
considers the critical theory perturbed by the combination $\lambda \Phi_p
+\tilde{\lambda} \tilde{\Phi}_p$~($\lambda >0$) \cite{ZAM1,MI}. In fact,
in the case of the minimal models, L\"assig \cite {MI}
 has identified such a phase for
$\tilde{\lambda} <0$, by using perturbative
RG calculations \cite{BZAM3,LUCA}. The generalization of this approach to all
A-D-E resonance models are technically complicated, because
the
necessary set of structure constants of the cosets models are not~(up to now)
explicitly known. However, we
point out that the field $\tilde{\Phi}_p$ has the correct conformal
dimensional to satisfy some of the arguments used in ref. \cite{MI}.
Moreover, at least for the ground state, our numerical results are
consistent with this picture.

Finally, we remark on some numerical observations concerning the
behaviour of the functions $L_a(\theta)$. In fig. 4 we show
this behaviour for the $A_3$ model. The Z(4)-symmetry assures that
$L_1(\theta)=L_3(\theta)$ and therefore we only plot the functions
$L_1(\theta)$ and $L_2(\theta)$ for $\theta >0$~($L_a(\theta)$ are
even functions). Roughly it seems that these functions differ only
by an overall constant~(for each ``p'') . This fact
 may be expected considering that
these functions are smooth and should assume constant values at the plateaux.
In
fig. 5 we have plotted the first derivative~(in the variable
$\theta$) of $L_1(\theta)$
and $L_2(\theta)$ for $X=-70$. We
believe that this result may motivate us to derive
more exact results from the TBA equation (10). Anyhow, for
the moment, this information has been used~(Appendix A) to estimate
the logarithmic correction to the Casimir energy in the ultraviolet limit.

\section{The ``$\phi^3$'' resonance S-matrix}
Let us consider a solution of Eq.(2) possessing a single particle $a$.
We also assume that the pole structure in the scattering amplitude produces
the particle $a$ itself.
Besides Eq.(2)
the S-matrix $S_{a,a}$ should satisfy the ``$\phi^3$'' bootstrap condition,
\EQ
S_{a,a}(\theta,\theta_0)= S_{a,a}(\theta +i \frac{\pi}{3},\theta_0)
S_{a,a}(\theta
-i \frac{\pi}{3},\theta_0)
\EN

Following the considerations of section 2 , the simplest solution of
Eqs. (2,17) are given by,
\EQ
S_{a,a}(\theta,\theta_0)=\frac{\tanh\frac{1}{2}(\theta+i\frac{2 \pi}{3})}
{\tanh\frac{1}{2}(\theta-i\frac{2 \pi}{3})}
\frac{\tanh\frac{1}{2}(\theta-\theta_0-i\frac{\pi}{3})}
{\tanh\frac{1}{2}(\theta-\theta_0+i\frac{\pi}{3})}
\frac{\tanh\frac{1}{2}(\theta+\theta_0-i\frac{\pi}{3})}
{\tanh\frac{1}{2}(\theta+\theta_0+i\frac{\pi}{3})}
\EN

The background part of $S_{a,a}(\theta,\theta_0)$ is the S-matrix \cite{YL1} of
the perturbed Yang-Lee edge singularity \cite{YL2}. The Toda related field
theory is that one analyzed by Arinshtein et al \cite{AFZ} and known as
the Shabat-Mikhailov model. This theory has only one field and the interaction
is through the potential $V(\phi)= e^{ \alpha \phi} +e^{-2 \alpha \phi}$,
$\alpha$ being the coupling constant. As
has been shown by Mikhailov \cite{MS},
this model is integrable and can be considered as a ``reduction'' of the
$A_2$ Toda field theory provided that the two fields $\phi_1,\phi_2$
in the $A_2$ model satisfy the relation
$\phi_1=-\phi_2=\phi$~(see ref. \cite{MS}). From the S-matrix
point of view, this can be seen by
defining the amplitude $S_{a,a}(\theta,\theta_0)$ as
\footnote{
A similar reduction as in Eq.(19) is possible
for any $A_{N-1}$ theory, $N $ odd. The next task is, however, the
identification of the reduced model for any kind of crossover behaviour
in non-unitary models.},
\EQ
S_{a,a}(\theta,\theta_0)= S_{1,1}(\theta,\theta_0)S_{1,2}(\theta,\theta_0)
\EN
where $S_{1,1}(\theta,\theta_0)$ and $S_{1,2}(\theta,\theta_0)$
are the $A_2$ amplitudes.

It follows from Eq.(19) and Eq.(10) that the pseudoenergie~($\epsilon(\theta)$)
of the model defined by Eq.(18) and those from the
$A_2$ Toda theory~($\epsilon_1(\theta),\epsilon_2(\theta)$) are equal,
namely $\epsilon(\theta)=\epsilon_1(\theta)=\epsilon_2(\theta)$. Hence
the
ground state energy associated to the S-matrix of Eq.(18) is precisely half
of that of the $A_2$ model analyzed in sect.3 .
Therefore,
the plateaux will now form around the values $c_p=1-12/p(p+1) ,p=4,5,..$ .
At this point it is important to recall
that the Casimir energy in non-unitary minimal models $M_{\frac{p}{q}}$
depends on both the
central charge and the lowest conformal dimension $\Delta_{min}$~(
effective central charge) \cite{ZU},
\EQ
E_0 \simeq -\frac{\pi}{6 R}(-24 \Delta_{min}+c) \simeq -\frac{\pi}{6R} c_{ef}
\EN
where  $c_{ef}=1-6/(pq)$.

This fact strongly suggests that we are dealing with non-unitary
minimal models. Indeed,
comparing $c_{ef}$ with $c_p=1-12/p(p+1), p=4,5,...$
we identify the following series of non-unitary minimal models:
$M_{\frac{q}{2q+1}}$~(q=p/2, p even) and
$M_{\frac{q+1}{2q+1}}$~(q=(p-1)/2, p odd).
We also recall
that for these models the field $\phi_{1,2}$ has the  minimal conformal
dimension. Motived by these analogies with the $A_2$ resonance
factorized theory it is also interesting to find the fields that may
drive the system to the staircase pattern of section 3.
{}From the expected finite size corrections to the ground
state we identify the fields $\phi_{2,1}$ and $\phi_{1,5}$ as those that
can induce these models to the crossover phenomenon. More precisely, we
expect the following behaviour for $M_{\frac{q+1}{2q+1}}$,
\EQ
M_{\frac{q+1}{2q+1}} + \phi_{2,1} \rightarrow M_{\frac{q}{2q+1}} ~~~q=2,3,...
\EN
and for $M_{\frac{q}{2q+1}}$,
\EQ
M_{\frac{q}{2q+1}} + \phi_{1,5} \rightarrow M_{\frac{(q-1)+1}{2(q-1)+1}}
{}~~~q=3,4,...
\EN

Moreover, the infrared regime is also governed by the operators $\phi_{2,1}$
and $\phi_{1,5}$.  An exception is the flow
$M_{3/7}+\phi_{1,5} \rightarrow M_{3/5}$ , where the field $\phi_{1,5}$
is not present in the $M_{3/5}$ Kac-table. In this case we think that
$\phi_{1,5}$ should be replaced by the level 2 descendent of $\phi_{1,3}$.
To the best of
our knowledge this pattern in which the fields $\phi_{2,1}$ and
$\phi_{1,5}$ interchange their roles as relevant and irrelevant
operators along the RG trajectory is new in the literature. In fig.(6)
we show this predicted behaviour \footnote{ It is notable
that the central charge increases only in the flow defined by Eq.(22).
Since these models are non-unitary they do not present any contradiction
with Zamolodchikov's c-theorem \cite{BZAM4,BZAM3}. For other interesting
behaviour in polymer systems see the discussion in ref. \cite{FESA} .}.

In order to give more support to this prediction we have numerically
studied
the spectrum of the Hamiltonian of the simplest flow,
namely $M_{3/5} +\phi_{2,1} \rightarrow M_{2/5}$. We have
constructed the Hamiltonian of this perturbed conformal field theory
on a torus of radius R using the truncated conformal approach
\cite{YZAM,MIMU,MIMUC}. In fig. (7) we show the evolution of the
first 40 eigenvalues as a function of the volume R. The $\phi_{2,1}$
perturbation allows us to divide the Hilbert space in two sectors
of the fields $[I,\phi_{2,1}]$~(dashed lines)
$[\phi_{1,2},\phi_{1,3}]$~(solid lines) and their descendents. The striking
feature is the absence of any crossing levels, that makes it difficult
to interpret fig.(7) as a consistent integrable massive theory. We recall
that in an integrable massive theory, there will be an abundance of level
crossings between states that are distinguished by their particle
content \cite{YZAM,MIMA}. For example, this will be the case for
momentum lines accumulating in the lowest particle threshold and
other n-particle states present in the theory. Moreover an a
posteriori analysis of the (possible) mass gaps are not consistent
(within our numerical precision) with, e.g., the exponential split
of the ground state. Although, we are not able to estimate the
infrared region using this approach,  fig.(7)
has a rather remarkable resemblance to typical massless crossover spectrum.
We believe that these observations bring some extra support to the
prediction of Eqs. (21,22). Finally, some additional comments are in order.

A first remark concerns a possible Ginsburg-Landau \cite{BZAM5}
interpretation of the simplest flow $M_{3/5} +\phi_{2,1}
\rightarrow M_{2/5}$. The fusion algebra for the operator $\phi_{1,2}$
with the
lowest dimension  in
the $M_{3/5}$ model is,
\EQ
\phi_{1,2} \otimes \phi_{1,2} \sim I +\phi_{1,3};~~~
\phi_{1,2} \otimes \phi_{1,3} \sim \phi_{2,1} +\phi_{1,2};~~~
\phi_{1,2} \otimes \phi_{2,1} \sim \phi_{1,3}
\EN

By identifying the operator $\phi_{1,2}$ with the elementary field $\Phi$
of the Ginsburg-Landau theory, it follows from Eq.(23) that $\phi_{1,3} \equiv
\Phi^2$ and $\phi_{2,1} \equiv \Phi^3$. The first descendent $L_{-1}
\bar{L}_{-1}
\phi_{1,2}$ being an odd field should be identified with $\Phi^5$. The
Ginsburg-Landau potential consistent with this last
identification is $V(\Phi)= \Phi^6$, which is in accordance with
the $Z(2)$ symmetry of $M_{3/5}$ model. Therefore, perturbing with
the field $\phi_{2,1}$ one can drive the $M_{3/5}$ model to a
$\Phi^3$ class of universality, which is precisely that of the
Yang-Lee edge singularity \cite{YL2} \footnote{ We also recall that
the identification of the operator $\phi_{1,3}$ is in reasonable
agreement with a $Z(2)$ spontaneously broken symmetry~(Ising-like)
found in its
spectrum \cite{MIMA}.}. It seems important to find the Ginsburg-Landau
picture of other non-unitary models. However, we have noticed that in
models with several negative and positive dimensions it seems quite
difficult to find a consistent description in
terms of only one basic field $\Phi$.

A second observation concerns the ambiguity of sign of the perturbing
coupling constant in the case of the $\phi_{1,5}$ operator. The
convention that is assumed in unitary models is that the positive
sign is related to the massless crossover and the negative sign to
a massive behaviour. Since these non-unitary models are
somewhat related ~(ground state) to an unitary theory ($A_2$ model)
one may expect that the same picture holds. However, in the
case of non-unitary theories we may also have phases in which
the system starts to present complex eigenvalues after
reaching a certain value
of R \cite{YZAM}. We believe that a further investigation is needed to
resolve this issue, which may end up in discovering new S-matrices
for non-unitary models.

\section{ Conclusions}
We have studied resonance factorized models based on the A-D-E Lie
algebra. By using the thermodynamic Bethe ansatz we have analyzed
the respective Casimir energy on a radius of length R. At
moderate distances of R we find a rich pattern of RG trajectories
associated with typical flows in the deformed GKO coset models.
We have pointed out a connection between the A-D-E resonance
scattering and one from the Toda field theory when its
coupling constant assumes special complex values. However,
the meaning of this analytical continuation in terms of a well
defined background quantum field theory is still to be found.
Moreover, it would be interesting
to investigate whether this is the only possible way
leading to factorized resonance S-matrices which can be interpreted
as a deformed conformal field theory.
We have obtained a general functional equation for the
functions $Y_a(\theta)=e^{\epsilon(\theta)}$ and one may
start with this equation instead of considering
the A-D-E resonance S-matrices. We recall that similar functional
equations appear in integrable lattice models
in the context of the
inversion relations to the transfer-matrix \cite{BA}. It seems
also that these equations hide a large amount of information
yet to be explored \cite{PEKU}.

The construction of a resonance factorized S-matrix satisfying
the ``$\phi^3$''-property has led us to predict new flows in
the non-unitary minimal models. In the case of the
simplest crossover, the spectrum presented
in section 4 may give us a guess on how the fields in the
ultraviolet regime will evolve to the infrared region. However,
in the more general cases considered in this paper, this property has
still to be disentangled.
It would  be also important to identify
the lattice models possessing such properties, the
$A_{2}^{2}$-RSOS models being a probable guess \cite{IK}.

\section*{Acknowledgements}
It is a pleasure to thank P. Haines and H.J.de Vega
for a careful reading of the manuscript and J.L. Cardy
for advise and encouragement.
\vspace{0.2cm}\\
{\bf Note added}: After we had submitted this paper to publication, we
received a copy of the preprint \cite{DR} by P. Dorey and F. Ravanini.
There is some overlap with the present paper concerning the A-D-E
generalization
of ref.[7].
\newpage
\section{Appendix A}
This appendix is concerned with the logarithmic corrections appearing
in Eq.(15). Our analysis is fairly parallel with that of ref.\cite{ZAM2}, in
order to generalize it to the case of many pseudoenergies. First we
notice that Eq.(7) permits us to write the expansion
\EQ
\frac{\tilde{\psi}_{a,b}}{2 \pi} =\sum_{n=0}^{\infty}
{(-1)}^{n} \frac{\tilde{\psi}_{a,b}^{(2n)}}{(2n)!} k^{2n}
\EN
where up to order $k^2$ we have,
\EQ
\tilde{\psi}_{a,b}^{(0)}= \delta_{a,b};~~~
\tilde{\psi}_{a,b}^{(2)}= 2{C^{-1}}_{a,b} \left ( \theta_0^2 +\frac{\pi^2}
{h^2} \right )
\EN

Using this expansion in Eq.(7), in the regime $X \rightarrow -\infty$,
we find
\EQ
m_a e^{\theta} +\ln(1-e^{-L_a(\theta)})=
\sum_{b} \sum_{n=1}^{\infty}
\frac{\tilde{\psi}_{a,b}^{(2n)}}{(2n)!} L_b^{(2n)}(\theta)
\EN
where $L_b^{(2n)}(\theta)= \frac{d^{2n}}{d^{2n} \theta} L_b(\theta)$.

Following ref. \cite{ZAM2}, the central charge in the region
$X \sim y$, $\theta -2X >>1$,
$y<<0$, satisfies the differential equation,
\EQ
\frac{\pi^2}{6} \frac{\partial}{\partial y}c(X,y) =-\sum_{a} m_a e^{y}L_a(y)
\EN
and one possible ansatz for its solution is given by,
\EQ
\frac{\pi^2}{6} =-\frac{1}{2} \sum_{a,b} \sum_{n=1}^{\infty}
\frac{\tilde{\psi}_{a,b}^{(2n)}}{(2n)!} \sum_{k=1}^{2n-1} {(-1)}^k
L_a^{(k)}(y) L_b^{(2n-y)}(y) \\
-\sum_{a} \int_{0}^{L_a(y)} \ln(1-e^{-t}) dt -\sum_{a} m_a e^{y} L_a(y)
\EN

The leading logarithmic correction is obtained by expanding
$\ln(1-e^{-t}) = -e^{-t} +...$, and by ignoring the term
$e^{\theta}$ in Eq.(26). The final result is,
\EQ
c(X,y)= r+ \frac{3}{2 {\pi}^2} \sum_{a,b}
\tilde{\psi}_{a,b}^{(2)} L_a^{(1)}L_b^{(1)}-\frac{6}{\pi^2}
\sum_{a} e^{-L_a(y)}
\EN
provided that the $L_a(\theta)$ satisfy,
\EQ
\sum_{b}\frac{\tilde{\psi}_{a,b}^{(2)}}{2} \frac{d^2}{d^2 \theta}
L_b(\theta) +e^{-L_a(\theta)}=0
\EN

Based on our numerical results, it is a reasonable approximation to consider
that the first derivatives of the $L_a(\theta)$ are index $a$ independent.
Therefore, we can have the following ansatz as a solution of Eq.(30),
\EQ
L_a(\theta)= \ln \left [ \frac{\sin^2 \lambda(\theta-\gamma)}
{\lambda^2 \sum_{b}\tilde{\psi}_{a,b}^{(2)}} \right ]
\EN
where $\lambda$ is fixed by periodicity to be $\lambda=\frac{\pi}{
2(X-\gamma_0)}$. $\gamma_0$ is a constant~(not determined by this approach)
such
that when $X \rightarrow \infty$, $\gamma  \rightarrow \gamma_0$. Substituting
Eq.(31) in Eq(29) we have,
\EQ
c=r- \frac{3}{2} \frac{\sum_{a,b} \tilde{\psi}_{a,b}^{(2)}}{X^2},~~
X \rightarrow \infty
\EN


\newpage
\centerline{\bf Figure Captions}
\vspace{0.5cm}
Fig. 1(a,b,c,d) The scaling function $c(R,\theta_0)$ for four values
of the resonance parameter $\theta_0$.
(a)$\theta_0=0$, (b) $\theta_0=10$, (c) $\theta_0=20$ and
(d) $\theta_0=40$.
\vspace{0.1cm}\\
Fig. 2(a,b,c) The beta function $\beta(g)$ for three values
of the resonance parameter $\alpha_0$. (a) $\theta_0=10$, (b) $\theta_0=20$
and (c) $\theta_0=40$.
\vspace{0.1cm}\\
Fig.3(a) The first RG plateaux for the models
$A_2$~(short dashed line), $A_3$~(solid line) and
$D_4$~(long dashed line).
\vspace{0.1cm}\\
Fig.3(b) The beta-function $\beta(g)$
for the models $A_2$~(short dashed line), $A_3$~(solid line) and
$D_4$~(long dashed line).
\vspace{0.1cm}\\
Fig.4 The functions $L_1(\theta)$~(solid line) and
$L_2(\theta)$~(dashed line) of the
$A_3$ model for $\theta_0=40$ and $X=-70$.
\vspace{0.1cm}\\
Fig.5 The first derivative of the
functions $L_1(\theta)$~(solid line) and $L_2(\theta)$~(dashed line) of the
$A_3$ model for $\theta_0=40$ and $X=-70$.
\vspace{0.1cm}\\
Fig.6 The flow pattern in the non-unitary minimal models
$M_{\frac{q}{2q+1}}; M_{\frac{q+1}{2q+1}},~~q=2,3,...$. The
horizontal(vertical)
arrows represent the relevant(irrelevant) operators defining the
ultraviolet(infrared) corrections to the fixed point.
\vspace{0.1cm}\\
Fig.7 The first 40 levels of the spectrum of the model $M_{3/5} +\phi_{2,1}$.
The dashed~(solid) lines correspond to the sector
 $[I,\phi_{2,1}]$~($[\phi_{1,2},\phi_{1,3}]$) and its descendents.

\begin{thebibliography}{99}
\bibitem{BPZ} A.A. Belavin, A.M. Polyakov, A.B. Zamolodchikov, {\em
Nucl.Phys.B 241 (1984) 333}
\bibitem{CAR} J.L. Cardy, {\em Phase Transitions and Critical Phenomena
(1987) vol 11 p 55, ed C Domb and J Lebowitz (New York, Academic)}
\bibitem{BZA1} A.B. Zamolodchikov, {\em JETP Letters 46 (1987) 160}
\bibitem{BZA2} A.B. Zamolodchikov, {\em Int.J.Mod.Phys. A 3 (1988) 743;
Advanced Studies in Pure Mathematics 19 (1989) 641}
\bibitem{ZAM1} Al.B. Zamolodchikov, {\em Nucl.Phys.B 342 (1990) 695}
\bibitem{YY} C.N. Yang, C.P. Yang, {\em J.Math.Phys.B 10 (1969) 1115}
\bibitem{ZAM2} Al.B. Zamolodchikov, {\em Paris preprint (1991) ENS-LPS-335}
\bibitem{BZAM3} A.B. Zamolodchikov,
{\em Yad.Fiz. 46 (1987) 82 [Sov.J.Nucl.Phys.
46 (1987) 1090]}
\bibitem{LUCA} A.W.W. Ludwig, J.L. Cardy, {\em Nucl.Phys.B 285 (1987) 687}
\bibitem{KMS} D.A. Kastor, E.J. Martinec, S.H. Shenker, {\em Nucl.Phys.B
316 (1989) 590}
\bibitem{MI1} M. L\"assig, {\em Phys.Lett.B 278 (1992) 439}\\
C. Ahn, {\em Cornell preprint (1992) CLNS 92/1135}
\bibitem{GKO} P. Goddard, A. Kent, D. Olive, {\em Commum.Math.Phys.
103 (1986) 105}
\bibitem{FLO} V.A. Fateev and S. Lukyanov,
{\em Kiev lectures (1988) IFT-88-74-76}\\
C. Crnkovic, G.M. Sotkov, M. Stanishkov, {\em Phys.Lett.B 226 (1989) 297}\\
M.T. Grisaru, A. Lerda, S. Penati, D. Zanon, {\em Nucl.Phys.B 346 (1990) 264}
\bibitem{MA} M.J. Martins, {\em SISSA preprint (1992) EP-72 }
\bibitem{KS} V. Kurak, J.A. Swieca, {\em Phys.Lett.B 82 (1979) 289}
\bibitem{AFZ} A.E. Arinshtein, V.A. Fateev, A.B. Zamolodchikov, {\em Phys.Lett.
B 87(1979) 389}
\bibitem{CRMU} P.Christe, G. Mussardo, {\em Nucl.Phys.B 330 (1990) 465;
Int.J.Mod.Phys. A5 (1990) 4581}
\bibitem{BC} H.W. Braden, E. Corrigan, P.E. Dorey, R. Sasaki,
{\em Phys.Lett.B 227 (1989)441; Nucl.Phys.B 338 (1990) 689}
\bibitem{DV} C. Destri, H.J.de Vega, {\em Phys.Lett.B 233 (1989) 336}
\bibitem{MU} G. Mussardo, {\em SISSA preprint (1992), to appear in Phys.Rep.}
\bibitem{ZAM3} Al.B. Zamolodchikov, {\em Phys.Lett.B 253 (1991) 391}
\bibitem{BCN} H.W.J. Blote,
J.L. Cardy, M.P. Nightingale, {\em Phys.Rev.Lett.56 (1986) 742}\\
I. Affleck, {\em Phys.Rev.Lett.56 (1986) 746}
\bibitem{WAL} V.A. Fateev, A.B. Zamolodchikov,
{\em Nucl.Phys.B 280 (1987) 694}\\
V.A. Fateev, S. Lukyanov, {\em Int.J.Mod.Phys.A 3 (1989) 507}
\bibitem{ZAMT} Al.B. Zamolodchikov, {\em Nucl.Phys.B 358 (1991) 497;
 358 (1991) 524; 366 (1991) 122}
\bibitem{TBA} V.A. Fateev, M.J. Martins, Al.B. Zamolodchikov, unpublished\\
M.J. Martins, {\em Phys.Lett.B 277 (1992) 301}\\
F. Ravanini, {\em Saclay preprint (1992) SPht/92-011}
\bibitem{MI} M. L\"assig, {\em Julich preprint (1991)}
\bibitem{YL1} J.L. Cardy, G. Mussardo, {\em Phys.Lett.B 225  (1989) 275}
\bibitem{YL2} M.E. Fischer, {\em Phys.Rev.Lett 40 (1978) 1610} \\
J.L. Cardy, {\em Phys.Rev.Lett 54 (1985) 1354}
\bibitem{MS} A.V. Mikhailov, {\em Pisma v ZhETF 30 (1979) 443 [JETP Letters
30 (1979) 414]}
\bibitem{ZU} C. Itzykson, H. Saleur, J.-B. Zuber, {\em Europhys.Lett.2 (1986)
91}
\bibitem{BZAM4} A.B. Zamolodchikov, {\em JETP Lett. 43 (1986) 730}
\bibitem{BZAM5} A.B. Zamolodchikov, {\em Yad. Fiz. 44 (1986) 82 [
Sov.J.Nucl.Phys. 44 (1986) 529]}
\bibitem{YZAM} Y.P Yurov, Al.B. Zamolodchikov, {\em Int.J.Mod.Phys.A 5
(1990) 3221; A6 (1991) 4557}
\bibitem{MIMU} M. L\"assig, G. Mussardo, {\em Computer.Phys.Comm. 66 (1991) 71}
\bibitem{MIMUC} M. L\"assig, G. Mussardo, J.L. Cardy, {\em Nucl.Phys.B 348
(1991) 591}
\bibitem{MIMA} M. L\"assig, M.J. Martins, {\em Nucl.Phys.B 354 (1991) 666}
\bibitem{FESA} P. Fendley, H. Saleur, {\em preprint (1992)
BUHEP-92-15/YCTP-P13}
\bibitem{BA} R.J. Baxter, {\em Exactly Solvable Models in Stastical
Mechanics~(Academic, London, 1982)}
\bibitem{PEKU} P.A. Pearce, A. Kl\"umper, {\em Phys.Rev.Lett. 66 (1991) 974}\\
A. Kl\"umper, P.A. Pearce, {\em J.Stat.Phys. 64 (1991) 13; Melbourne preprint
N0 23 (1991)}\\
P.A. Pearce, {\em Melboune preprint N0 22 (1991)}
\bibitem{IK} A.G. Izergin, V.E. Korepin, {\em Commum.Math.Phys. 79 (1981)
303}\\
A. Kuniba, {\em Nucl.Phys.B 355 (1991) 801}
\bibitem{DR} P.E. Dorey, F. Ravanini, {\em Saclay/Bologna preprint SPhT-92-065/
DFUB-92-09}
\end{thebibliography}
\end{document}